\documentstyle[epsfig,longtable]{aipproc}

\renewcommand{\[}{\begin{equation}}
\renewcommand{\]}{\end{equation}}

\def\epl#1#2#3{Europhys. Lett. {\bf #1}, #2 (19#3)}

\def\prb#1#2#3{Phys. Rev. B {\bf #1}, #2 (19#3)}
\def\prl#1#2#3{Phys. Rev. Lett. {\bf #1}, #2 (19#3)}

\begin{document}

\setlength{\unitlength}{1cm}
\noindent \begin{picture}(0,0)
\put(0,2){\noindent \sf Presented at the:}
\put(0,1.5){\noindent \sf ``Fifth Williamsburg Workshop on
First--Principles Calculations for Ferroelectrics'',}
\put(0,1){\noindent \sf February 1--4, Williamsburg, VA, USA}
\end{picture}
\title{The quantum-mechanical position operator and the
polarization problem}

\author{Raffaele Resta}
\address{INFM -- Dipartimento di Fisica Teorica, \\
Universit\`a di Trieste, Strada Costiera 11, 
I-34014 Trieste, Italy \\ and \\
Department of Physics, The Catholic
University of America, Washington, D.C. 20064}

\maketitle

\begin{abstract} The position operator (defined within Schr\"odinger
representation as usual) becomes meaningless when the usual Born-von
K\'arm\'an periodic boundary conditions are adopted: this fact is at the
root of the polarization problem.  I show how to define the position
expectation value by means of rather peculiar many--body (multiplicative)
operator acting on the wavefunction of the extended system. This
definition can be regarded as the generalization of a precursor work,
apparently unrelated to the polarization problem. For uncorrelated
electrons, the present finding coincides with the so--called
``single--point Berry phase'' formula, which can hardly be regarded as the
approximation of a continuum integral, and is computationally very useful
for disordered systems. Simulations which are based on this concept are
being performed by several groups.  \end{abstract}

\section*{Introduction}

In a phenomenological description of dielectric media the concept of
macroscopic polarization is the basic one \cite{Landau}. One formally
defines the polarization of a macroscopic sample as its electrical dipole
moment, divided by the volume. The problem is that the dipole of a finite
system is dominated by surface effects, which seem to make polarization
ill defined as a property of the bulk. In condensed matter theory, one
typically gets rid of undesirable surface effects by dealing with extended
systems within Born-von K\'arm\'an periodic boundary conditions (BvK).
Unfortunately, this does not solve the polarization problem: the dipole is
in fact the expectation value of the quantum--mechanical electronic
position, which in its usual form is an ill defined operator within BvK.

Because of the above reasons, even {\it defining} what bulk dielectric
polarization is remained a major challenge for many years. If one assumes
the polarization elements as discrete, \`a la Clausius-Mossotti, then the
polarization mechanism can be safely understood: but such oversimplified
picture does not apply to a real dielectric, where the electronic
distribution is continuous, and often very delocalized. Typically,
textbooks attempt to explain polarization of a bulk periodic solid via the
dipole moment of a unit cell, or something of the  kind
\cite{Kittel,Ashcroft}.  These definitions are incorrect \cite{Martin74}:
according to the modern viewpoint, bulk macroscopic polarization is a
physical observable {\it completely independent} from the periodic charge
distribution of the polarized dielectric.

The breakthrough was fostered by the 1992 Williamsburg meeting. Macroscopic
polarization was {\it defined} in terms of the wavefunctions, not of the
charge. This definition has an unambiguous thermodynamic limit, such that
BvK and Bloch states can be used with no harm~\cite{rap73}. In the
following months a modern theory of macroscopic polarization in crystalline
dielectrics has been completely established \cite{rap78,rap_a12}, thanks to
a major advance due to R.D. King-Smith and D. Vanderbilt \cite{King93}, who
expressed polarization in terms of a Berry phase \cite{Berry84,Berry}. An
early comprehensive account of the modern theory exists \cite{rap_a12}. 
Other less technical presentations are available as well
\cite{rap82,Martin97a}; for an oversimplified outline see
Ref.~\cite{rap_a18}.

In its original form, the Berry--phase theory of polarization was based on
independent--electron wavefunctions, such as the Kohn--Sham orbitals of
density--functional theory \cite{rap_a12,DFT}. Its generalization to
correlated many--electron wavefunctions is due to Ort\'{\i}z and Martin
\cite{Ortiz94}. Both the independent--electron and the
correlated--electron versions of the Berry--phase theory of polarization
rely on some lattice periodicity, and define macroscopic polarization in
the form of a reciprocal--space integral. In particular, the
Ort\'{\i}z--Martin polarization of a correlated many--electron system
takes the form of a peculiar ``ensemble average'', since it is an integral
over a set of many--body ground states. This viewpoint is indeed correct
and very valuable: the Ort\'{\i}z--Martin theory has even been implemented
to study the polarization of interesting model systems of highly
correlated electrons \cite{rap87,Ortiz95,rap90}. However, the fact that
macroscopic polarization could not be even {\it defined} by means a ``pure
state'' expectation value (over a single many-electron ground state) was a
disturbing drawback. This has been overcome thanks to the advance of
Ref.~\cite{rap100}, where a novel solution to the polarization problem is
provided. The compact formula arrived at in Ref.~\cite{rap100},
Eq.~(\ref{limit}) below, is {\it apparently} unrelated to the Berry--phase
concept; it applies on the same footing to correlated systems and to
independent--electron systems, as well as to crystalline and to disordered
systems. Lattice periodicity and integration in reciprocal space are no
longer needed in order to {\it define} what polarization is.  Polarization
can in fact be cast as the expectation value of a rather peculiar operator
over the many--electron ground wavefunction (in the thermodynamic limit).

In the present work, I present the main achievement of Ref.~\cite{rap100}
under a rather different light than in the original paper. I will show
that its main result can be regarded as the many--electron generalization
of a precursor work, apparently unrelated to the polarization problem
\cite{Selloni87}. I will also discuss in detail the independent--electron
case, linking in particular the results of Ref.~\cite{rap100} to the
so--called single--point Berry phase\cite{rap_a17,rap101}, a concept
which proves very useful, particularly in the study of disordered systems.

\section*{The Hilbert space of condensed matter}

Almost invariably, in condensed matter theory one adopts BvK for the
electronic wavefunctions, when dealing with either crystalline systems or
disordered systems. There are several good reasons for such choice. One
reason is that one eliminates any surface by construction, thus getting
rid of indesirable surface effects. Another reason is that in the
crystalline case one can exploit the virtues of Bloch's theorem. I stress
that the BvK choice is tantamount to defining the Hilbert space where our
solutions of Schr\"odinger equation live.

For the sake of simplicity, the present work will deal with the
one--dimensional case. For a single--particle wavefunction BvK reads
$\psi(x+L) = \psi(x)$, where $L$ is the imposed periodicity, chosen to be
large with respect to atomic dimensions.  Notice that lattice periodicity
is {\it not} assumed, and BvK applies to disordered systems as well.  An
operator maps any vector of the given Hilbert space into another vector
belonging to the same space: the multiplicative position operator $x$ is
{\it not} a legitimate operator when BvK are adopted for the state
vectors, since $x \,\psi(x)$ is not a periodic function whenever $\psi(x)$
is such. Of course, any periodic function of $x$ is a legitimate
multiplicative operator: this is the case {\it e.g.} of the nuclear
potential acting on the electrons.

In the following, we will need to explicitly introduce many--body
wavefunctions as well. BvK then imposes periodicity in each electronic
variable separately: \begin{equation} \Psi_0(x_1, \dots, x_i, \dots, x_N)
= \Psi_0(x_1, \dots, x_i\! + \! L, \dots, x_N) .  \label{perio}
\end{equation} Our interest is indeed in studying a bulk system: $N$
electrons in a segment of length $L$. Eventually the thermodynamic limit
is taken: $L \rightarrow \infty$, $N \rightarrow \infty$, and $N/L = n_0$
constant.  We will also assume throughout the ground state nondegenerate,
and we deal with insulating systems only: this means that the gap between
the ground eigenvalue and the excited ones remains finite for $L
\rightarrow \infty$.  

In order to illustrate where the main problem is, let us start with an
Hilbert space {\it different} from the BvK one, using instead the boundary
conditions which are appropriate for the bound states of an isolated atom
or molecule. In this case the many--body wavefunction $\Phi_0(x_1, \dots,
x_i, \dots, x_N)$ goes to zero exponentially whenever $|x_i|$ gets large.
Within this Hilbert space there is no problem with the position, which is
trivially defined as the multiplicative operator \begin{equation} \hat{X} =
\sum_{i=1}^N x_i \label{X} \end{equation} within the Schr\"odinger
representation. The ground--state expectation value is then simply:
\begin{equation} \langle X \rangle = \langle \Phi_0 | \hat{X} | \Phi_0
\rangle = \int \!dx \; x \, n(x), \label{trivial} \end{equation} where
$n(x)$ is the one--particle density. The value of $\langle X \rangle$
scales with the system size, and the quantity of interest is indeed the
dipole per unit length, which coincides with macroscopic polarization.  The
expectation value in Eq.~(\ref{trivial}) is dominated by surface effects,
thus making the definition of a bulk quantity problematic.  If instead one
adopts BvK,  Eq.~(\ref{perio}), then the operator defined by Eq.~(\ref{X})
is no longer a legitimate one, as explained above.

The major goal is therefore to define the expectation value of the
electronic position $\langle X \rangle$ in the BvK Hilbert space, and to
prove that our definition provides in the thermodynamic limit the physical
macroscopic polarization of the sample: this has been very recently
achieved in Ref.~\cite{rap100}. Before illustrating this major advance, we
discuss an important precursor work, apparently unrelated with the
polarization problem, where nonetheless the expectation value of the
position operator plays the key role.

\section*{The electron--in--broth formula}

\begin{figure}[b] \begin{center} \setlength{\unitlength}{1cm}
\begin{picture}(9.2,4.5) \put(0,0.3){\includegraphics{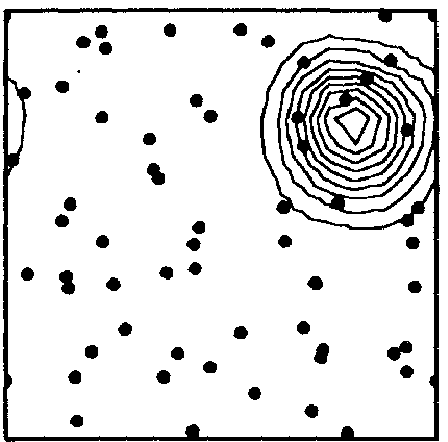}}
\put(5,0.3){\includegraphics{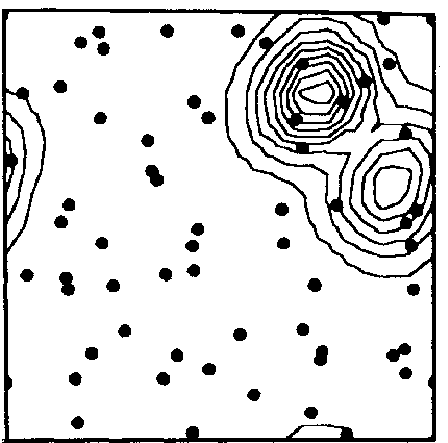}}  \end{picture} \caption{Contour
plot of the electronic density (integrated along the sight line)  at two
different time steps. The dots are projections of the ionic positions. 
After Ref.~\protect\onlinecite{Selloni87}.} \end{center} \end{figure}

Some years ago, A. Selloni {\it et al.} \cite{Selloni87} addressed the
properties of electrons dissolved in molten salts at high dilution, in a
paper which at the time was commonly nicknamed the ``electron in broth''. 
The physical problem was studied by means of a mixed quantum--classical
simulation, where a lone electron was adiabatically moving in a molten
salt (the ``broth'') at finite temperature. The simulation cell contained
32 cations, 31 anions, and a single electron. KCl was the original case
study, which therefore addressed the liquid state analogue of an F center;
other systems were studied afterwards \cite{brodo}. The motion of the
ions was assumed as completely classical, and the Newton equation of
motions were integrated by means of standard molecular dynamics (MD)
techniques, though the ionic motion was coupled to the quantum degree of
freedom of the electron. The electronic ground wavefunction was determined
solving the time--dependent Schr\"odinger equation at each MD time step. 
Two snapshots of the simulation, reproduced from the original paper, are
shown in Fig.~1.

As usual in MD simulations, periodic boundary conditions are adopted for
the classical ionic motion. Ideally, the ionic motion occurs in a
simulation cell which is surrounded by periodic replicas: inter-cell
interactions are accounted for, thus avoiding surface effects.
Analogously, the electronic wavefunction is chosen
in the work of Selloni {\it et al.} to obey BvK over the simulation cell,
and therefore features periodic replicas as well. In Fig.~1, the ``center''
of the electron distribution is close to one of the borders of the
simulation cell: the contour plot clearly shows the tail of its replica
spilling into the opposite border.

One of the main properties investigated in Ref.~\cite{Selloni87} was the
electronic diffusion, where the thermal ionic motion is the driving agent
(within the adiabatic approximation). In order to perform this study, one
has to identify first of all where the ``center'' of the electronic
distribution is.  Intuitively, the distributions in Fig.~1 appear to have
a ``center'', which however is defined only modulo the replica
periodicity, and furthermore {\it cannot} be evaluated  simply as $\int d
{\bf r} \, {\bf r} |\psi({\bf r})|^2$ precisely because of BvK. Selloni
{\it et al.} solved the problem by means of a very elegant and
far--reaching formula.  Here I write its one--dimensional analogue in
the form: \begin{equation} \langle x \rangle = \frac{L}{2\pi} \mbox{Im log
} \int_0^L \!\! dx \;  {\rm e}^{i\frac{2\pi}{L} x} |\psi(x)|^2 .
\label{one} \end{equation} This formula provides the center of the
electronic distribution, which is defined unambiguously though modulo $L$,
as it must be.

\section*{How the formula works}

The density of the single--particle wavefunction $n(x) = |\psi(x)|^2$ obeys
BvK periodicity and can therefore be written as a Fourier series:
\begin{equation} n(x) = \sum_{s = -\infty}^\infty C_s \, {\rm
e}^{i\frac{2\pi s}{L}x}. \label{four} \end{equation} The
electron--in--broth formula, Eq.~(\ref{one}) becomes: \begin{equation}
\langle x \rangle = \frac{L}{2\pi} \mbox{Im log } \int_0^L \!\! dx \;  {\rm
e}^{i\frac{2\pi}{L} x} n(x) = \frac{L}{2\pi} \mbox{Im log } C_{-1},
\label{brodo} \end{equation} a quantity which is well defined (modulo $L$)
whenever the $s\!=\!-1$ Fourier coefficient is nonvanishing.

In general, any periodic function $n$ can be written---in a nonunique
way---as the superposition of a localized function $n_{\rm loc}$ and of its
periodic replicas. We write in the most general case the superposition as
\begin{equation} n(x) = \sum_{n = -\infty}^\infty n_{\rm loc} (x - x_0 -nL)
, \end{equation} where $x_0$ is hitherto arbitrary.  Then the Fourier
coefficients of $n(x)$ are related to the Fourier transform of $n_{\rm
loc}(x)$ as: \begin{equation} C_s = \frac{1}{L} \tilde{n}_{\rm loc}
(\frac{2\pi s}{L}) \, {\rm e}^{- i\frac{2\pi s}{L}x_0}. \end{equation} We
therefore express $\langle x \rangle$, Eq.~(\ref{one}), as:
\begin{equation} \langle x \rangle = \frac{L}{2\pi} \mbox{Im log } \left[
{\rm e}^{i\frac{2\pi}{L} x_0} \tilde{n}_{\rm loc} (- \frac{2\pi}{L})
\right] = x_0 + \frac{L}{2\pi} \mbox{Im log } \tilde{n}_{\rm loc} (-
\frac{2\pi}{L}) . \label{brodo2} \end{equation} 

We now specialize to the only interesting case: $n_{\rm loc}(x)$ is a real
positive function, which integrates to one over $(-\infty, \infty)$;
furthermore $n_{\rm loc}(x)$ is centered at $x\!=\!0$ and decays over
distances of the order of $\Delta x$. In the special case where $n_{\rm
loc}(x)$ is an even function, its Fourier transform is real and
Eq.~(\ref{brodo2}) simplifies to $\langle x \rangle \equiv x_0$ (mod $L$),
which is indeed the expected result. In the general case where $n_{\rm
loc}(x)$ is {\it not} an even function, an expansion of its Fourier
transform is needed. The Fourier transform $\tilde{n}_{\rm loc}(k)$ decays
over a reciprocal distance of the order of $1/\Delta x$: therefore
supposing $\Delta x \ll L$ (as {\it e.g.} is clearly the case in
Fig.~1) the expansion yields: \begin{equation} \tilde{n}_{\rm loc} (-
\frac{2\pi}{L}) = \int_{-\infty}^\infty \!\! dx \, n_{\rm loc}(x) + i
\frac{2\pi}{L} \int_{-\infty}^\infty \!\! dx \, x n_{\rm loc}(x) + {\cal
O}(L^{-2}) .  \end{equation} Since by hypothesis $n_{\rm loc}(x)$
integrates to one, Eq~(\ref{brodo2}) to leading order yields:
\begin{equation} \langle x \rangle \simeq  x_0 + \int_{-\infty}^\infty
\!\! dx \, x n_{\rm loc}(x) \qquad (\mbox{mod} \; L), \end{equation} again
a very meaningful expression for the center of the electronic distribution
$n(x)$.

\section*{Polarization as a many--electron problem}

In the previous two Sections we have discussed how to define the
expectation value of the position operator within BvK, when only a single
electron is present in a large cell. When dealing instead with
macroscopic polarization, we are faced with a system of many electrons,
having a finite density. As anticipated above, our main interest is in
studying a bulk system: $N$ electrons in a segment of length $L$.
Eventually the thermodynamic limit is taken: $L \rightarrow \infty$, $N
\rightarrow \infty$, and $N/L = n_0$ constant. If one arrives at defining
a suitable position expectation value $\langle X \rangle$, analogue of
Eq.~(\ref{trivial}) for the much less trivial BvK case, then the
electronic contribution to macroscopic polarization (dipole per unit
length) is: \begin{equation} P_{\rm el} = \lim_{L \rightarrow \infty}
\frac{e \langle X \rangle}{L} , \label{limit2} \end{equation} where $e$
is the electron charge.

So the main issue is as follows: can the electron--in--broth formula,
Eq.~(\ref{one}), be generalized to provide an expectation value $\langle X
\rangle$ appropriate to the case of  many electrons in the given
BvK periodicity? The answer is yes, but a smart generalization is needed.
The most straightforward generalization, where one simply identifies $n(x)$
with the electron density of the extended system, does not work at all.
There is a very good reason for this.  

Suppose the system is crystalline, and furthermore assume independent
electrons for the sake of symplicity. According to the modern
viewpoint~\cite{rap_a12} the macroscopic polarization is a bulk observable
{\it completely independent} from the periodic charge distribution of the
polarized dielectric. While the density is obtained from the square modulus
of the Bloch orbitals, polarization can only be obtained from their {\it
phase}. Indeed, if one attempts to actually evaluating Eq.~(\ref{one}) upon
replacement of $|\psi(x)|^2$ with the bulk density $n(x)$ of the extended
system a meaningless result is reached. In fact, if the lattice constant of
the crystalline solid is $a$, the BvK periodicity is then taken over a
multiple of $a$: $L\!=\!Ma$. Since $n(x)$ is periodic with period $a$, its
Fourier coefficients $C_s$ in Eq.~(\ref{four}) vanish unless $s$ is a
multiple of $M$: therefore $C_{-1}\!=\!0$, and the electron--in--broth
formula, Eq.~(\ref{brodo}), is ill defined.

In order to arrive at the correct many--electron generalization of
Eq.~(\ref{one}), one first introduces the multiplicative many--body
operator ${\rm e}^{i\frac{2\pi}{L} \hat{X}}$, where $\hat{X}$ is the
simple operator defined in Eq.~(\ref{X}). In terms of this, one then
defines: \begin{equation} \langle X \rangle = \frac{L}{2\pi} \mbox{Im log
} \langle \Psi_0 | {\rm e}^{i\frac{2\pi}{L} \hat{X}} | \Psi_0 \rangle . 
\label{main} \end{equation} The similarity with Eq.~(\ref{one}) is
self evident; as in the one--body case, the position expectation value is
defined only modulo $L$. The main ingredient of our generalized
electron--in--broth formula is a many--body expectation value: an
$N$--electron integral where a genuine {\it many--body operator} appears. 
In general, one defines an operator to be one--body whenever it is the
{\it sum} of $N$ identical operators, acting on each electronic coordinate
separately: for instance, the $\hat{X}$ operator itself is such. In order
to express the expectation value of a one--body operator the full
many--body wavefunction is not needed: knowledge of the reduced one--body
reduced density matrix $\rho$ is enough: this is {\it e.g.} the case in
Eq.~(\ref{trivial}). I stress that, instead, the expectation value of
${\rm e}^{i\frac{2\pi}{L} \hat{X}}$ over a correlated wavefunction {\it
cannot} be expressed in terms of $\rho$, and knowledge of the $N$-electron
wavefunction is explicitly needed. The uncorrelated case where $| \Psi_0
\rangle$ is a single determinant will be discussed in the next Section.

The electronic polarization, Eq.~(\ref{limit2}), becomes then:
\begin{equation} P_{\rm el} = \lim_{L \rightarrow \infty} \frac{e}{2\pi}
\mbox{Im log } \langle \Psi_0 | {\rm e}^{i\frac{2\pi}{L} \hat{X}} | \Psi_0
\rangle. \label{limit} \end{equation} As anticipated in the Introduction,
this compact and general expression for the macroscopic polarization
applies on the same footing to correlated systems and to
independent--electron systems, as well as to crystalline and to disordered
systems.  Notice that in any case $L \rightarrow \infty$ is an highly
nontrivial limit, since the exponential operator in Eq.~(\ref{limit}) goes
formally to the identity, but the size of the system and the number of
electrons in the wavefunction increase with $L$.  Therefore some {\it
ad--hoc} procedure is needed to demonstrate Eq.~(\ref{limit}): this is
reported in detail in Ref.~\onlinecite{rap100}, and will not be repeated
here. 

\section*{Independent particles and single--point Berry phase}

We focus here on an uncorrelated system of independent electrons, whose
$N$-electron wavefunction $| \Psi_0 \rangle$ is a Slater determinant. In
this special case $| \Psi_0 \rangle$ is uniquely determined by the
one--body reduced density matrix $\rho$ (which is the projector over
the set of the occupied spinorbitals): therefore the expectation
value $\langle X \rangle$, Eq.~(\ref{main}), is uniquely determined by
$\rho$. The expression for $\langle X \rangle$---as well as the related
one for the electronic polarization---is however most simply expressed
directly in terms of the orbitals, as I am going to show. Such an
expression, which goes under the name of ``single--point Berry phase'',
was first proposed in a series of lecture notes~\cite{rap_a17}; it has
been further scrutinized in some detail in a recent paper by Yaschenko
{\it et al.} \cite{rap101}.

Suppose $N$ is even, and $| \Psi_0 \rangle$ is a singlet.  The Slater
determinant has thus the form: \begin{equation} | \Psi_0 \rangle =
\frac{1}{\sqrt{N!}} |\varphi_1 \overline{\varphi}_1 \varphi_2
\overline{\varphi}_2 \dots \varphi_{N/2} \overline{\varphi}_{N/2}| ,
\label{slater} \end{equation} where $\varphi_i$ are the single-particle
orbitals. It is then expedient to define \begin{equation} | \tilde{\Psi}_0
\rangle =  {\rm e}^{i\frac{2\pi}{L} \hat{X}} | \Psi_0 \rangle :
\label{tilde} \end{equation} even $| \tilde{\Psi}_0 \rangle$ is indeed a
Slater determinant, where each orbital $\varphi_i(x)$ of $| \Psi_0 \rangle$
is multiplied by the plane wave $ {\rm e}^{i\frac{2\pi}{L} x}$. According
to a well known theorem, the overlap amongst two determinants is equal to
the determinant of the overlap matrix amongst the orbitals. We therefore
define the matrix (of size $N/2\times N/2$): \begin{equation} S_{ij} =
\int_0^L \! dx \, \varphi^*_i(x) {\rm e}^{i\frac{2\pi}{L} x} \varphi_j(x),
\end{equation} in terms of which we easily get \begin{equation} \langle X
\rangle = \frac{L}{2\pi} \mbox{Im log } \langle \Psi_0 | \tilde{\Psi}_0
\rangle = \frac{L}{\pi} \mbox{Im log det } S , \label{single}
\end{equation} where the factor of 2 accounts for spin.

Notice that nowhere we have assumed crystalline symmetry: the $\varphi_i$
orbitals will therefore have the most general form, while obviously
obeying BvK at the sample boundary. Eq.~(\ref{single}) applies on the same
ground to crystalline and to disordered systems: in the crystalline case
it can be shown to be equivalent to a discretization of the usual
continuum Berry phase~\cite{rap_a12}, where crystalline Bloch orbitals are
used. A simple test case is studied by Yaschenko {\it et al.}
\cite{rap101}, where the numerical equivalence is actually demonstrated.

The single--point Berry phase has been recently implemented quite
successfully to calculate infrared spectra of disordered
systems~\cite{simulations,Pasquarello97}. In a typical Car--Parrinello
simulation~\cite{CP}, the thermodynamic limit is approximated using BvK
over a large simulation cell, which contains several tens of atoms.  The
time evolution of the ions is followed along discrete time steps, while
the electrons follow adiabatically. In order to evaluate the infrared
spectrum, one needs to evaluate the electronic macroscopic current
traversing the simulation cell. This is done on the fly with the use of
the single--point Berry phase, where $\langle X \rangle$,
Eq.~(\ref{single}), is evaluated at each time step, and its time
derivative is approximated with a finite difference~\cite{simulations}.  

\section*{Conclusions}

The formula originally proposed by Selloni {\it et al.} \cite{Selloni87}
for dealing with a lone electron in a large liquid sample, and named
here the ``electron--in--broth formula'', is quite naturally generalized to
the many--body case. It provides then a novel solution to the polarization
problem, Eq.~(\ref{limit}): such solution, though apparently unrelated to
the Berry--phase concept, is indeed a discretized Berry phase in
disguise~\cite{rap100}.

Since any Berry phase is by definition a phase, it may appear surprising
that in Eq.~(\ref{limit}) only the square modulus of the wavefunction
enters: any phase information is then apparently obliterated.  One has to
bear in mind, however, that $| \Psi_0 \rangle$ is the {\it many--body}
wavefunction of the $N$--electron system. In the independent--electron
case, this wavefunction embeds the relevant information about the
relative {\it phases} of the one--electron orbitals, and this is what one
needs in order to evaluate the macroscopic polarization; in the
correlated--electron case, the wavefunction carries in a more complex way
the analogue information.

\section*{Acknowledgments}

Discussions with M. Bernasconi and E. Yaschenko are gratefully
acknowledged.  Work partly supported by the Office of Naval Research,
through grant N00014-96-1-0689.

\end{document}